\documentclass[aps,prb,twocolumn,showpacs,superscriptaddress,groupedaddress]{revtex4-1}  
\usepackage{graphicx}  
\usepackage{dcolumn}   
\usepackage{bm}            
\usepackage{amsmath}   
\usepackage{amssymb}   
\usepackage[latin1]{inputenc}
\usepackage{tikz}
\usetikzlibrary{shapes,arrows}
\usepackage{mathrsfs}
\usepackage{fixfoot,xspace}
\DeclareFixedFootnote*{\repnote}{See Supplemental Material for the form of the potential.} 

\begin{document}

\title{Origin of static and dynamic steps in exact Kohn-Sham potentials}

\author{M.\ J.\ P.\ Hodgson, J.\ D.\ Ramsden and R.\ W.\ Godby}
\affiliation{Department of Physics, University of York and European Theoretical Spectroscopy Facility, Heslington, York YO10 5DD, United Kingdom}
\date{\today}
\begin{abstract}
Knowledge of exact properties of the exchange-correlation (xc) functional is important for improving the approximations made within density functional theory. Features such as steps in the exact xc potential are known to be necessary for yielding accurate densities, yet little is understood regarding their shape, magnitude and location. We use systems of a few electrons, where the exact electron density is known, to demonstrate general properties of steps. We find that steps occur at points in the electron density where there is a change in the `local effective ionization energy' of the electrons. We provide practical arguments, based on the electron density, for determining the position, shape and height of steps for ground-state systems, and extend the concepts to time-dependent systems. These arguments are intended to inform the development of approximate functionals, such as the mixed localization potential (MLP), which already demonstrate their capability to produce steps in the Kohn-Sham potential.
\end{abstract}

\pacs{31.15.E-, 71.15.Mb, 31.15.A, 73.63.-b}

\maketitle
\section{Introduction}

Density-functional theory\cite{PhysRev.136.B864} (DFT) and time-dependent DFT\cite{PhysRevLett.52.997,PhysRevLett.82.3863} have been applied widely to calculate the properties of ground-state and time-dependent systems of interacting electrons.
In some cases the approximations made in practice perform extremely well; in others they become less valid, and hence the accuracy of the approach suffers. While the Kohn-Sham\cite{PhysRev.140.A1133} (KS) formulation of DFT is in principle exact, the scope of practical DFT calculations is limited by our understanding of the exact exchange-correlation (xc) potential. Therefore identifying important features that are missing from the common approximations, and developing new approximations which incorporate these features, is crucial.

Steps in the xc potential (a jump in the level of the xc potential over a relatively short distance) have been shown to be crucial for an accurate description of the electron density for a variety of ground-state and time-dependent systems\cite{almbladh1985density,vanLeeuwen_steps,:/content/aip/journal/jcp/131/22/10.1063/1.3271392,PhysRevA.54.1957,PhysRevLett.109.036402,PhysRevLett.109.266404,PhysRevA.85.022514,PhysRevB.88.241102,PhysRevA.88.042508,PhysRevB.90.241107,PhysRevA.90.042501}, such as tunneling electrons and charge transfer/excitations. Atomic structure calculations by van Leeuwen \textit{et al.}\cite{vanLeeuwen_steps} demonstrated that steps arise at the boundaries between atomic shells. Yang \textit{et al.}\cite{PhysRevA.90.042501}, using ensemble DFT, showed how, as more atomic KS orbitals are occupied, steps form in the exact xc potential. However, much remains to be understood regarding their position, shape and magnitude. 

Common approximate functionals struggle to model systems such as those above, as well as molecular dissociation, Van der Waals interaction and open-shell molecules\cite{:/content/aip/journal/jcp/122/9/10.1063/1.1858371}. Therefore improved functionals must be developed, thus understanding features, such as steps in the xc potential, is of great importance. 

We study the nature of steps that form in the KS potential for asymmetric ground-state and time-dependent, `molecule-like' systems (where the external potential tends to zero far from any atom), and expand the concept to symmetric systems. We examine the precise shape, height and position of steps, and show how steps combine to make other features in $v_\mathrm{xc}$, even in the time-dependent regime.

In Section~\ref{Almbladh-von Barth} we begin our analysis by considering the thought experiment of Almbladh and von Barth\cite{almbladh1985density}, where a step in the xc potential forms for a finite system of two spin-$\tfrac{1}{2}$ electrons. By analyzing the effect of the step on the electron density, we deduce the principles underlying the position, height and shape of steps, applying even when multiple KS orbitals are occupied. We then extend these ideas to the time-dependent regime. We derive, from these principles, arguments for the position and magnitude of steps, to aid the development of approximate functionals which have the ability to produce steps in $v_\mathrm{xc}$, such as the mixed localization potential (MLP)\cite{PhysRevB.90.241107}.

In Sections~\ref{The origin of steps}-\ref{effect of delocalization} we model finite systems in one dimension using our iDEA code\cite{PhysRevB.88.241102} in which we find the exact xc potential by first solving the time-dependent many-electron Schr{\"o}dinger equation to obtain the fully correlated wavefunction. From this we calculate the \textit{exact electron density} for ground-state, and subsequently time-dependent, systems. We then reverse-engineer the KS potential via an optimization algorithm which matches the non-interacting density to the interacting density. Our main calculations use \textit{spinless} electrons in order to explore systems with more correlation for a given computational cost, i.e.\ with each electron occupying a different KS orbital. Our focus will be on nano-wires and devices for which one-dimensional descriptions are appropriate, and hence we use the appropriately softened Coulomb repulsion $\left (  \left | x'-x \right |+ 1 \right )^{-1}$ (in atomic units). 

\section{The Almbladh-von Barth thought experiment}
\label{Almbladh-von Barth}

When using DFT to simulate neutral molecules, such as that described below, the use of local and semilocal density functionals to approximate the xc potential gives rise to errors that affect observables such as binding energy curves and energy surfaces. These errors arise in part due to the inability of such approximations to correctly predict the amount of charge on each atomic site\cite{:/content/aip/journal/jcp/125/19/10.1063/1.2387954,perdew1982density,perdew1985density}, therefore it is essential for the development of improved functionals to understand the role of the xc step in accurately localizing the KS electrons within the molecule.

\begin{figure}[htbp]
  \centering
  \includegraphics[width=1.0\linewidth]{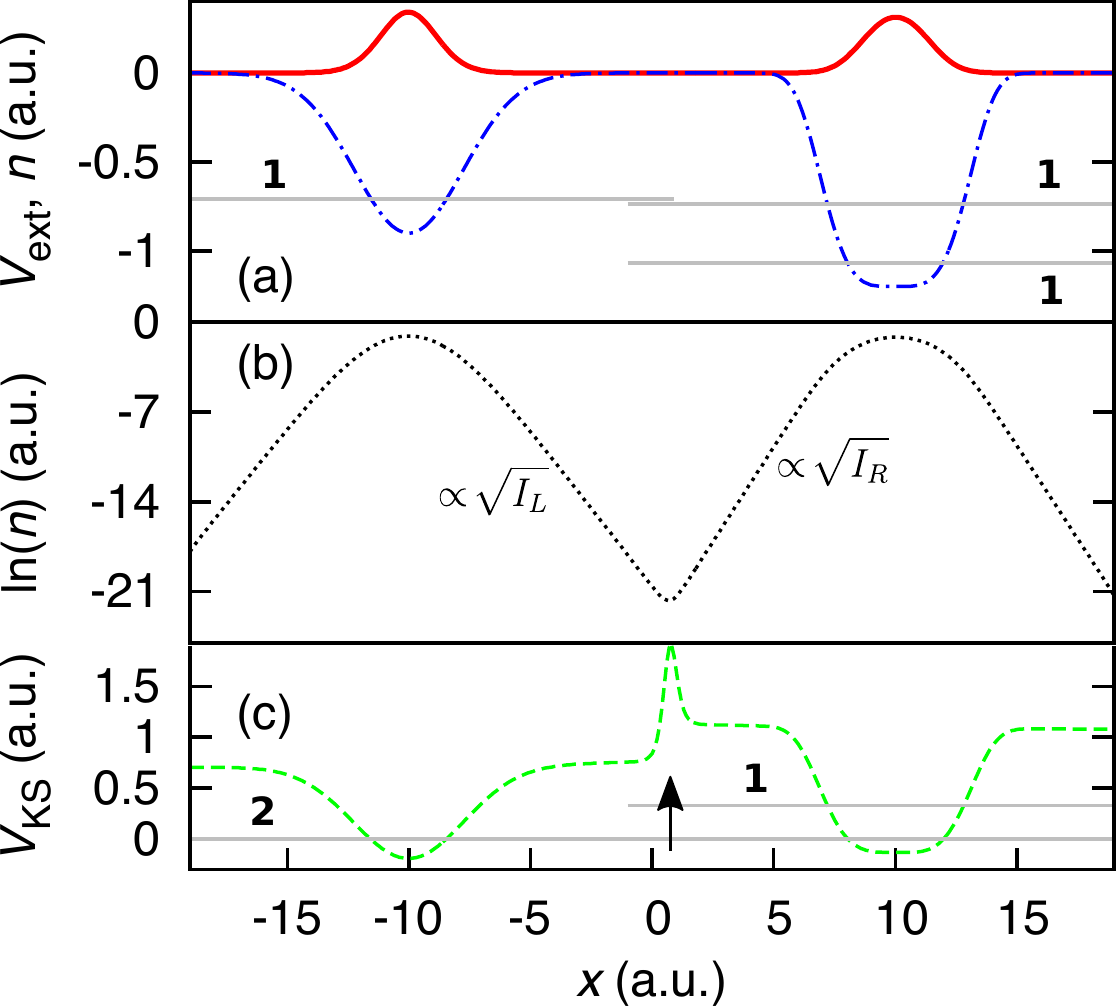}
\caption{(color online) (Two spin-$\tfrac{1}{2}$ electrons in two separated wells) --- (a) The external potential (dotted-dashed blue), together with the electron density for two \textit{interacting, spin-$\tfrac{1}{2}$} electrons (solid red). The horizontal gray lines show the bound single-particle energy states of the potential and the number adjoining each energy level indicates the degeneracy of that state. (b) The natural logarithm of the density, allowing the density minimum to be clearly identified. The decay of $\ln{(n)}$ on either side of the density minimum is proportional to the square root of the ionization energy of the well the electron occupies. (c) The exact KS potential (dashed green): the step of height $I_R-I_L$ (arrow) ensures that one electron is in each well. Note that the step aligns the ground-state energies of the two wells, as anticipated by Almbladh and von Barth\cite{almbladh1985density}.}
\label{Almbladh-von_Barth}
\end{figure}

We consider a one-dimensional double-well external potential\repnote, representing two separated open-shell atoms, where the right well has single-particle energy states that are lower than those of the left well\footnote{See Supplemental Material for details on all systems}. Owing to their Coulomb repulsion, two \textit{spin-$\tfrac{1}{2}$, interacting} electrons occupy different wells; however two \textit{noninteracting} electrons would both occupy the right-hand well; see Fig~\ref{Almbladh-von_Barth}(a). Hence, a step must form in the KS potential to allow the KS electron density to match the many-body density. This system has been studied for many years, originally by Almbladh and von Barth\cite{almbladh1985density} and Perdew\cite{perdew1985density}.

If we consider each individual well separately, as a subsystem, then the ground-state energies are equal to minus the ionization energies of the respective atoms\cite{PhysRevLett.51.1884} (wells) since $v_\mathrm{ext}(\left | x \right |\rightarrow \infty)=0$. ($I_R$ represents the ionization energy of the right well and $I_L$ is that of the left well. Considering the left and right atoms as individual systems, or subsystems, is valid for well separated atoms, and in the disassociation limit the concept is exact.)

Treated individually, both subsystem's KS potential decays to different, but approximately spatially constant, values, therefore at their region of intersection in the complete system a step exists whose height is the difference between those constants. 
We define the step height as $S_\mathrm{xc} \equiv v^\mathrm{xc}_R-v^\mathrm{xc}_L = v^\mathrm{KS}_R-v^\mathrm{KS}_L$, where $v^\mathrm{KS}_R$ and $v^\mathrm{KS}_L$ are the constants of the KS potentials in the right and left subsystems respectively, and likewise for the xc potentials $v^\mathrm{xc}_R$ and $v^\mathrm{xc}_L$. This definition is exact in the limit that the wells are infinitely separated, as the xc potential tends to a constant value far from the subsystem, hence the step acts to shift $v_\mathrm{xc}$ by $S_\mathrm{xc}$ between the subsystems. We find, however, that the formula holds well for electrons with only a few \r{a}ngstr\"{o}ms of separation; see Fig.~\ref{Almbladh-von_Barth}(c).

We reverse-engineer the exact KS potential\footnote{Reverse-engineered potentials, such as this one, are determined to within an additive constant.} for this system and, as predicted\cite{almbladh1985density,perdew1985density}, we observe a step in the xc potential between the wells; see Fig.~\ref{Almbladh-von_Barth}(c). The argument made by Almbladh and von Barth was that the step must align the KS single-particle energy levels of the two wells in order for the highest occupied molecular orbital (HOMO) to have sufficient weight in each well, i.e.\ one electron's worth of charge per atom (well). Hence, the step must have a magnitude which equals the difference between the HOMO energies of the two wells. 

While the above argument is robust for this system, we may come to the same conclusion via a different point of view. Consider now the form of the electron density far from any atom. Even for the many-body case, the density will decay asymptotically like that of a single particle occupying the well\cite{perdew1982density,PhysRevB.56.16021} $n(x) \propto e^{-2\sqrt{2I}x}$. As only one KS orbital is occupied for this system, the single orbital approximation (SOA)\cite{PhysRevB.90.241107,smith1979density} is exact (up to an additive constant). Applying the SOA to the density of Fig.~\ref{Almbladh-von_Barth}(a), we find that \textit{at the density minimum} the xc potential jumps by $I_R-I_L$; see Fig.~\ref{Almbladh-von_Barth}(b) and (c). The SOA is correctly sensitive to the decay of the electron density either side of the step when the density is of the form $e^{-2\sqrt{2I}x}$, a result also observed by Helbig \textit{et al.}\cite{:/content/aip/journal/jcp/131/22/10.1063/1.3271392}. Thus, at the interface between the electrons, where the density decaying from the left meets the density decaying from the right (the density minimum), the potential jumps from $I_L$ to $I_R$; see Fig.~\ref{Almbladh-von_Barth}(b). As this happens over a short range, a sharp step forms. Therefore the step can be considered to arise from this change in the decay of the electron density, which we will hence forth refer to as a change in the \textit{`local effective ionization energy'}. 

Below we study systems where more than one KS orbital is occupied. We find that a change in the local effective ionization energy remains responsible for steps. However, owing to the analog of this effect in the KS picture (see Section~\ref{Position_of_steps}), the magnitude and shape of steps can change. 

\section{The origin of steps}
\label{The origin of steps}

To begin this Section, we detail why, in general, the magnitude of the step may change for systems with more than one occupied KS orbital. In the following section we explore the effect the magnitude of the step has on the electron density, and whether the step height $I_R-I_L$ (given by the SOA in all cases) is a good approximation for the step height in a general system. 

We consider the form of the decay of the density either side of the step for both the many-body picture and the KS picture, in order to more fully understand what determines the step's magnitude in general. In the many-body picture, the density decaying from the left-hand subsystem (more generally -- simply decaying from the left), as the wells are separated far from one another, is given by $n'_L(x)  \propto e^{-2\sqrt{2I_L}x}$. Likewise, the right-hand subsystem contributes $n'_R(x)  \propto e^{+2\sqrt{2I_R}x}$. The decay of the density coming from the left-hand subsystem in the KS picture is $n_L(x)  \propto e^{ -2\sqrt{2\left(v^\mathrm{KS}_L-\varepsilon_L\right)}x }$, where $\varepsilon_L$ is the energy of the highest occupied KS orbital that dominates the asymptotic density of the left-hand subsystem. And for the right-hand subsystem $n_R(x) \propto e^{ +2\sqrt{2\left(v^\mathrm{KS}_R-\varepsilon_R\right)}x }$, where $\varepsilon_R$ is defined correspondingly. As $n'=n$, for the exact KS potential, it must follow that $v^\mathrm{KS}_L=I_L+\varepsilon_L$ and $v^\mathrm{KS}_R=I_R+\varepsilon_R$ (within an overall additive constant). Noting that the step height is $S_\mathrm{xc}= v^\mathrm{KS}_R-v^\mathrm{KS}_L$, it follows that
\begin{equation}
S_\mathrm{xc} = (I_R-I_L)+(\varepsilon_R-\varepsilon_L),
\label{step_height}
\end{equation}
where a negative value indicates a step that drops when going from left to right, and a positive value \textit{vice versa}. Equation~\ref{step_height} is exact in the limit that the atoms are infinitely separated, however, we have found the equation to be accurate for separations of a few \r{a}ngstr\"{o}ms. (Eq.~\ref{step_height} requires knowledge of the exact KS eigenenergies $\varepsilon_R$ and $\varepsilon_L$, determined partially by the step, and hence cannot be used to \textit{predict} the step height.)

The energies $\varepsilon_L$ and $\varepsilon_R$ refer to the highest occupied KS orbitals that dominate the density in the outer region of each subsystem. When the system consists of localized, well-separated subsystems, this concept is well defined, and it is in this case that a sharp step may form in $v_\mathrm{KS}$. Where the subsystems are closer and the electrons less localized, we find that the energies remain a useful interpretive concept.

Equation~\ref{step_height} shows that the step arises from two effects: the change in the local effective ionization energy in the many-body picture ($I_R-I_L$), and its counterpart in the KS picture ($\varepsilon_R-\varepsilon_L$); see Section~\ref{Position_of_steps} later.
Thus, the overall step can be considered as the sum of two steps, $S_\mathrm{xc} = S^I_{\mathrm{xc}} + S^{\varepsilon}_{\mathrm{xc}}$, where $S^I_{\mathrm{xc}} = I_R - I_L$ and $S^{\varepsilon}_{\mathrm{xc}} = \varepsilon_R - \varepsilon_L$; see Section~\ref{superposition}.

The above argument applies to spin-$\tfrac{1}{2}$ electrons as well as spinless electrons. We note that if we apply the above logic to a system consisting of spin-$\tfrac{1}{2}$ electrons, where there is an \textit{odd} number of electrons on each site, the highest occupied KS orbital must be spread over both wells. Hence, in this case $\varepsilon_R=\varepsilon_L$, and therefore the step height is that of the Almbladh-von Barth thought experiment discussed above ($S_\mathrm{xc} = I_R - I_L$).

When developing approximate xc functionals, there are certain known exact properties that one aims to satisfy, such as the derivative discontinuity of the xc energy with respect to electron number\cite{perdew1982density}. The derivative discontinuity predicts a jump in the xc potential by a constant as the electron number passes through an integer, which may lead one to connect steps in the xc potential with the derivative discontinuity. However, it is apparent from the above analysis that the magnitude of the step in $v_\mathrm{xc}$ is a result of the precise way in which the electron density decays from each subsystem. The decay of the density, in the many-body picture, and in the KS picture, has no direct association with the electron affinity of the subsystem ($-\varepsilon_{N+1}(N+1)$), nor the affinity of the system as a whole. We therefore conclude that the step which forms in $v_\mathrm{xc}$ is not attributed to the derivative discontinuity. For example, in the Almbladh-von Barth system this insensitivity to the affinity is complete.

\section{Height of steps}

Next we examine the effect that under- or overestimating the step height would have on the electron density. For example, noting that any step given by the SOA is of height $I_R-I_L$, we may ask whether this is an appropriate value for the step height in a general system. To answer this, one must consider the effect that altering the step height has on the electron density. 

For spin-$\tfrac{1}{2}$ electrons in a separated double-well system where the occupied KS orbitals are atomic orbitals, changing the step height equates to adding a constant to the potential within a given subsystem, and so usually affects the density only in the region of the step (see below). However, if the change in step is too large -- enough to alter the occupation of the wells -- the electron density will be affected everywhere. In the case where the highest occupied orbital is spread over both sites, the step height must be exactly $I_R-I_L$. 

Building on the arguments of Perdew\cite{perdew1985density} for the range of allowed energies of a system connected to a reservoir, we find a range for our step height for our molecular system consisting of spinless electrons. If we consider a system where, in the KS picture, $M$ states are filled in the left well, and $N$ states are filled in the right well, we can place a range on the magnitude of the step that must exist in $v_\mathrm{xc}$ based on correctly filling the eigenstates of the individual wells; see Fig.~\ref{step_range}. For this case we are assuming that the wells are sufficiently separated so that the single-particle eigenenergies ($\varepsilon$) of each well are unaffected by the electrons in the other well, other than the shift by a constant due to the step -- in all cases this degree of well separation would be needed in order for a sharp step to form. 

\begin{figure}[htbp]
  \centering
  \includegraphics[width=1.0\linewidth]{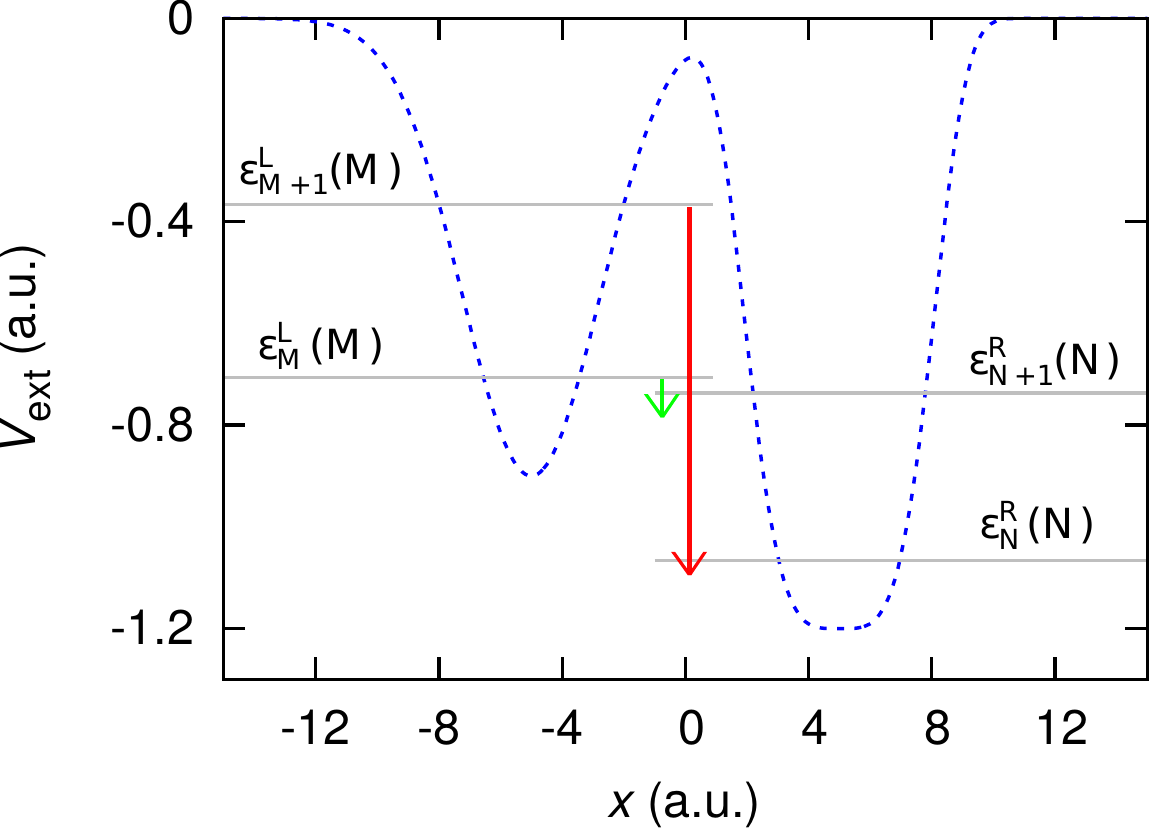}
\caption{(color online) The external potential (dashed blue) for the a general double-well molecule,  the lines indicate the bound single-particle energy levels of the individual wells, where, in this case, $M=N=1$. As it is, the external potential in the absence of interaction would give the incorrect filling of each well, i.e.\ both electrons in the right hand well. The green arrow indicates the minimum step height to achieve one electron per well, and the (longer) red arrow indicates a step height that is too large. Any value of $S_\mathrm{xc}$ between the two values would give a fairly accurate electron density.}
\label{Step_height_range}
\end{figure}

We know that the HOMO of the left well ($\varepsilon^L_M(M)$), plus the shift in energy due to the step ($S_\mathrm{xc}$, without loss of generality we set $v^\mathrm{KS}_R = 0$), must be less than the lowest unoccupied molecular orbital (LUMO) of the right well ($\varepsilon^R_{N+1}(N)$), and vice versa, allowing the amount of charge in each well to be correct. Thus, we can infer that 
\begin{equation}
\varepsilon^L_M(M)-\varepsilon^R_{N+1}(N)<S_\mathrm{xc}<\varepsilon^L_{M+1}(M)-\varepsilon^R_N(N). 
\label{step_range}
\end{equation}
A schematic representation of the range is shown in Fig.~\ref{Step_height_range}. The external potential has been chosen such that the lowest two single-particle states are located in the right-hand well, thus in the KS picture the step must correct this to allow the lowest two energy states of the overall system to be located in separate wells. The green arrow indicates the minimum the step height can be, whereas the red (long) arrow shows a step that is too large. These limits define the allowed range for the step height, in order for the electron density to be reasonably accurate.

Equation~\ref{step_range} applies also for spin-$\tfrac{1}{2}$ electrons (noting that the number of electrons will be different, as two electrons may occupy each energy level), except for the case where there is an odd number of electrons on each site. In this case the above arguments do not apply, however the step height is known exactly ($S_\mathrm{xc} = I_R-I_L$; derived above). 

Finally, we look at how changes in the step height affect the detailed electron density in the region of the step, and hence show which features of the density determine the exact step height. Consider a finite molecule that is very similar to the Almbladh and von Barth thought experiment (System 1), except two KS states are now occupied as opposed to one; two \textit{spinless} electrons, where the external potential is a double well\repnote, designed such that the first excited state of the right-hand well is lower than the ground state of the left-hand well (Fig.~\ref{step_range}). Hence, in the absence of interaction, both spinless electrons would occupy the lowest \textit{two} states of the right-hand well. As the many-electron density has one electron's worth of charge in each well due to the Coulomb repulsion and Pauli exchange, the exact KS potential must form a step to achieve this in the KS density. The step acts to shift the ionization energy of the two wells here, as opposed to aligning them, allowing the ground-state of the left-hand well to be lower in energy than the first excited state of the right-hand well, in accordance with our range (Fig.~\ref{step_range}). 

\begin{figure}[htbp]
  \centering
  \includegraphics[width=1.0\linewidth]{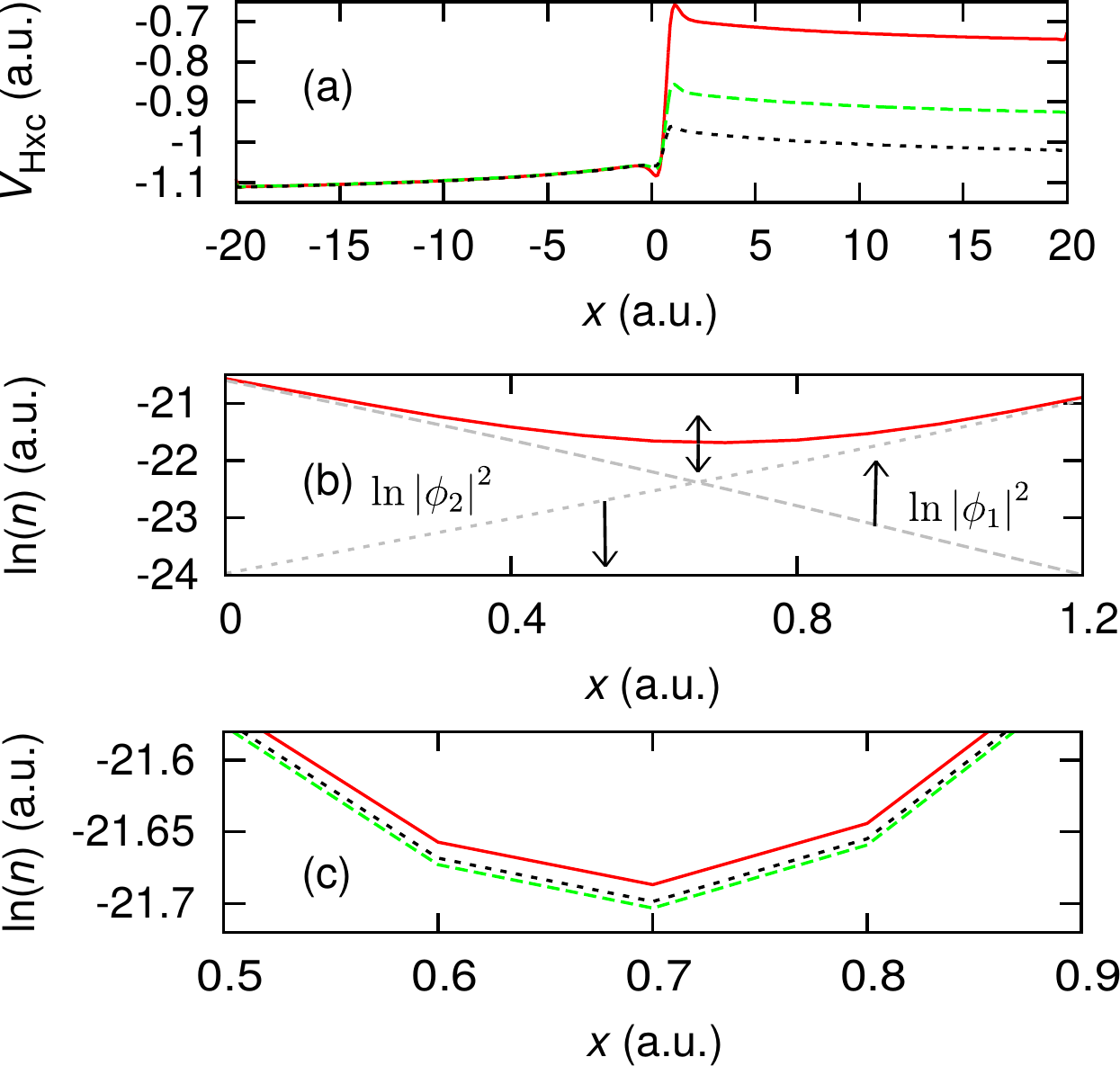}
\caption{(color online) System 1 (two spinless electrons in an asymmetric double well) --- (a) The exact $v_\mathrm{Hxc}$ potential (solid red) and two artificial stepped $v_\mathrm{Hxc}$ potentials (long-dashed green and dashed blue). (b) Natural log of the density at the density minimum. The natural log of the KS densities for the ground-state ($\left | \phi_1 \right |^2$) and the first excited state ($\left | \phi_2 \right |^2$) are shown. As the step height is decreased these densities change (indicated by the arrows), and thus affect the overall density by increasing or decreasing the magnitude at the density minimum (determined by the precise way the KS densities superimpose). (c) The densities corresponding to the step heights of (a), where the colors and line styles correspond.}
\label{Step_height_den}
\end{figure}

Figure~\ref{Step_height_den} shows how an artificially imposed change in step height affects the electron density. We observe that the change to the electron density is small, provided the step height is in the range given by Eq.~\ref{step_range}. The change in step height has the effect of reducing or increasing the density minimum very slightly. Precisely how the density minimum is affected is determined by the individual KS densities, i.e.\ $n_1 = \left | \phi_1 \right |^2$ and $n_2 = \left | \phi_2 \right |^2$. As the magnitude of the step is decreased, less of the right-hand KS density tunnels through to the left, and the opposite effect happens for the left-hand KS density. Thus we can conclude that \textit{the step height ultimately determines the degree to which the left-hand electron occupies the right well and vice versa} -- this applies to all cases. Thus, local and semilocal approximations to the xc potential must be exceedingly sensitive to changes in the density at the location of the step, or else a fully nonlocal approximation must be employed.

\section{Position of steps}
\label{Position_of_steps}

We consider the xc potential far from a molecule (i.e.\ several subsystems), hence the subsystems are no longer distinguishable. And therefore, the density must decay with the ionization energy of the \textit{whole} molecule, which in the case of a molecule comprised of separated atoms is the \textit{lowest} ionization energy all the wells. This means that for any subsystem's density which does not decay with the ionization energy of the whole system, there must be a \textit{second} change in the local effective ionization energy far from the system, and therefore another step must form. This second step was first observed by Perdew\cite{perdew1985density} and also by Makmal \textit{et al.}\cite{PhysRevA.83.062512} in the exact exchange potential for LiF, where they attribute the steps to shifts in the KS eigenvalues. They discuss the `domain' of each atom being dominated by the HOMO of that atom, resulting in a plateau to correct for the non-zero asymptotic limit caused by the HOMO eigenvalue being non-zero. This is the analog of the change in the local effective ionization energy in the KS picture. Hence, generally, this causes a step in the exact KS potential in accordance with our derivation of Eq.~\ref{step_height}. Thus, the `overall' step in the exact xc potential is a combination of the steps caused by the change in local effective ionization energy and the crossover of the single-particle KS densities (see below). When correlation effects are taken into account \textit{both} these effects must also be considered.

We define, as a function of space, $\tilde{I}(x) =\frac{1}{8n^2}\left ( \frac{\partial n}{\partial x} \right )^2$ (which is the second term in the SOA\@ expression for the KS potential [Eq.~1 in Ref.~\onlinecite{PhysRevB.90.241107}], hence showing the correct sensitivity of the SOA to the ionization energy), which represents the local effective ionization energy when the density decays asymptotically, which is true for regions of the density near the edge of a subsystem. Hence, in such a region $\tilde{I}(x)=I$, and between subsystems $\tilde{I}(x)$ may have a step demonstrating the change in the local effective ionization energy. (While in this paper we apply this formula to spinless electrons, the concept applies to spin-$\tfrac{1}{2}$ electrons also.)

\begin{figure}[htbp]
  \centering
  \includegraphics[width=1.0\linewidth]{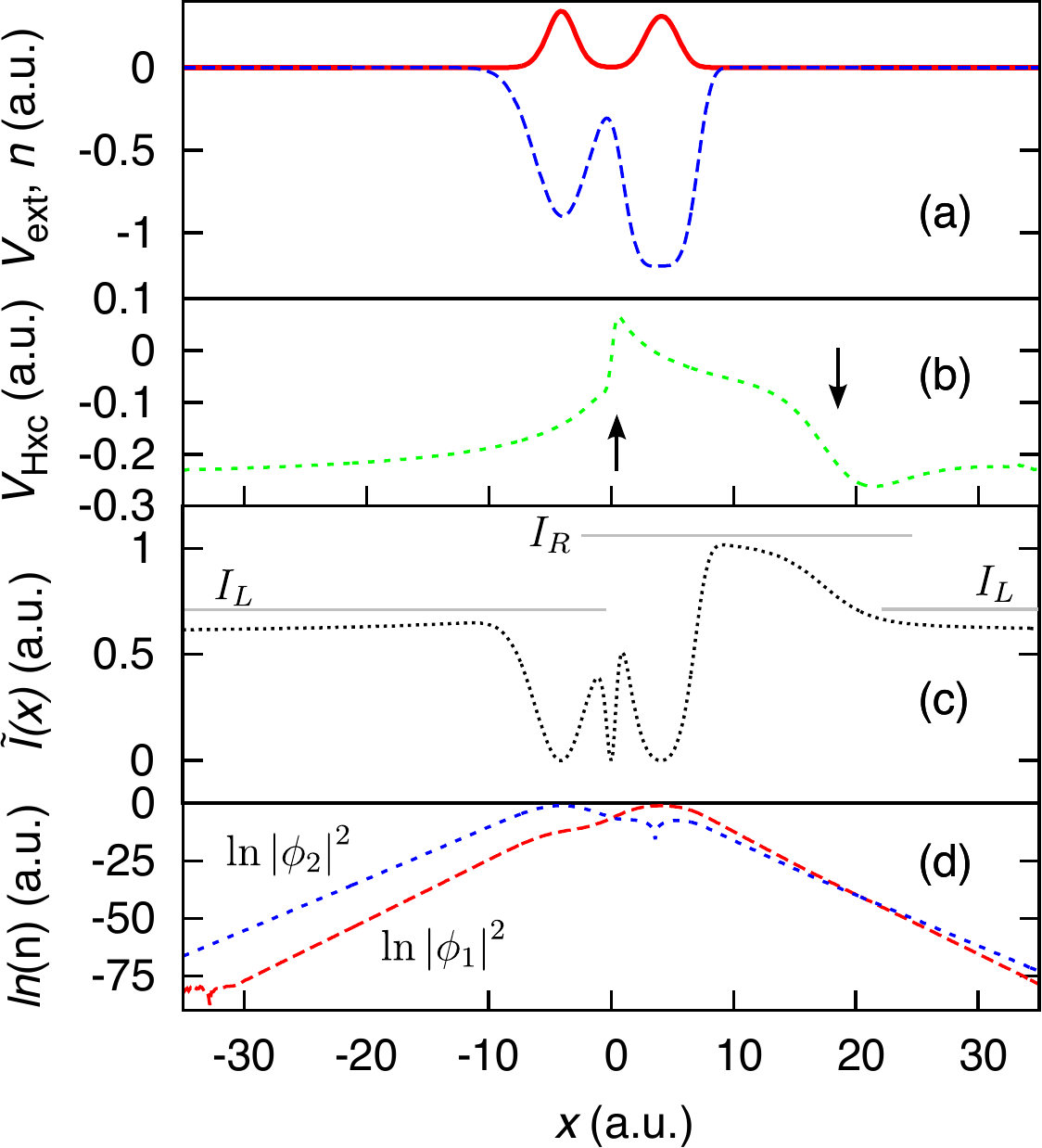}
\caption{(color online) System 2 (two spinless electrons in a molecule) --- (a) The electron density (solid red) and external potential (dashed blue). (b) The Hxc potential (dashed green), the arrows indicate where the steps are. One step forms at the density minimum, where the KS single-particle densities cross. The second step forms far from the molecule, again where the KS densities cross. (c) There are two changes in the local effective ionization energy (black dotted) in the many-body density, each corresponding to a step in $v_\mathrm{xc}$, the gray lines indicate the ionization energies of the subsystems ($I_L$ and $I_R$). (d) The natural log of the KS densities, blue short-dashed is the first excited KS density and red dashed is the ground-state KS density. As the decay rate of the first excited-state must be less than that of the ground-state, far from the molecule the densities must cross.}
\label{Extended}
\end{figure}

Figure~\ref{Extended}(a) shows a molecular system (System 2)\repnote where we observe the second, postulated step far from the molecule; see Fig.~\ref{Extended}(b). In Ref.~\onlinecite{PhysRevA.83.062512} the correcting step is observed for the exact exchange potential. Our step is, in part the same as this correcting step, however it superimposes with the a step which forms as a result of a change in local effective ionization energy at the same point in space. Also in Ref.~\onlinecite{:/content/aip/journal/jcp/131/22/10.1063/1.3271392} the second step was deduced to exist, however, was not observed. Our findings show that their thinking was correct, as our argument here applies to the spin-$\tfrac{1}{2}$ case (as well as for our spinless electrons). Furthermore, Fig.~\ref{Extended}(c) and (d) show that the step forms at the point where there is a crossover of the single-particle KS densities, i.e.\ where the dominant contributing single-particle density switches (applying also to spin-$\tfrac{1}{2}$ electrons for systems where more than one KS orbital is occupied). This is consistent with the findings of Ref.~\onlinecite{PhysRevA.83.062512} (discussed above) and Ref.~\onlinecite{vanLeeuwen_steps}, where the xc potential has ``a clear step structure and is constant within the atomic shell and changes rapidly at the atomic shell boundaries'' (also where the local ionization energy can change). Reference~\onlinecite{PhysRevA.90.042501} also found that a step structure forms when more than one orbital begins to be occupied.

Here we observe that the change in the local effective ionization energy and the crossover in the KS single-particle densities manifest at the same point, hence the two steps superimpose ($S_\mathrm{xc}=S^I_\mathrm{xc}+S^\varepsilon_\mathrm{xc}$). In general, a change in the dominant single-particle KS density corresponds to a change in the local effective ionization energy, but not necessarily vice versa. For example, in the Almbladh-von Barth system there is a change in the local effective ionization energy without a crossover of the localized KS densities, since only one orbital is occupied ($\varepsilon_R=\varepsilon_L \Rightarrow S_\mathrm{xc}=I_R-I_L$). 

Della Sala and G{\"o}rling showed that for a three dimensional system, along a direction $\textbf{r}$ which corresponds to a nodal surface of the HOMO, the exact xc potential will approach a non-zero constant\cite{PhysRevLett.89.033003}. Our analysis can be generalized to 3D, and agrees with this result; if the HOMO is zero in the direction $\textbf{r}$, then, as $r \rightarrow \infty$, the dominant contribution to the overall density from the single-particle KS densities must come from the highest occupied KS orbital that does not correspond to a nodal surface. Hence, the non-zero KS density and the `true' HOMO KS density cannot cross. Thus the counteracting step we observe in Fig.~\ref{Extended} will not manifest and the xc potential may tend to a non-zero constant.

The role of the KS orbitals in this argument is reminiscent of the appearance of KS orbitals in meta-GGA\cite{PhysRevLett.91.146401,PBE} functionals and the Becke-Edgecombe electron localization function (ELF)\cite{:/content/aip/journal/jcp/92/9/10.1063/1.458517}, and draws attention to the power of the KS orbitals in improving density functionals. Our MLP approximation, likewise, makes use of the KS orbitals in defining the degree of localization. 

\subsection{Time-dependence}

\begin{figure}[htbp]
  \centering
  \includegraphics[width=1.0\linewidth]{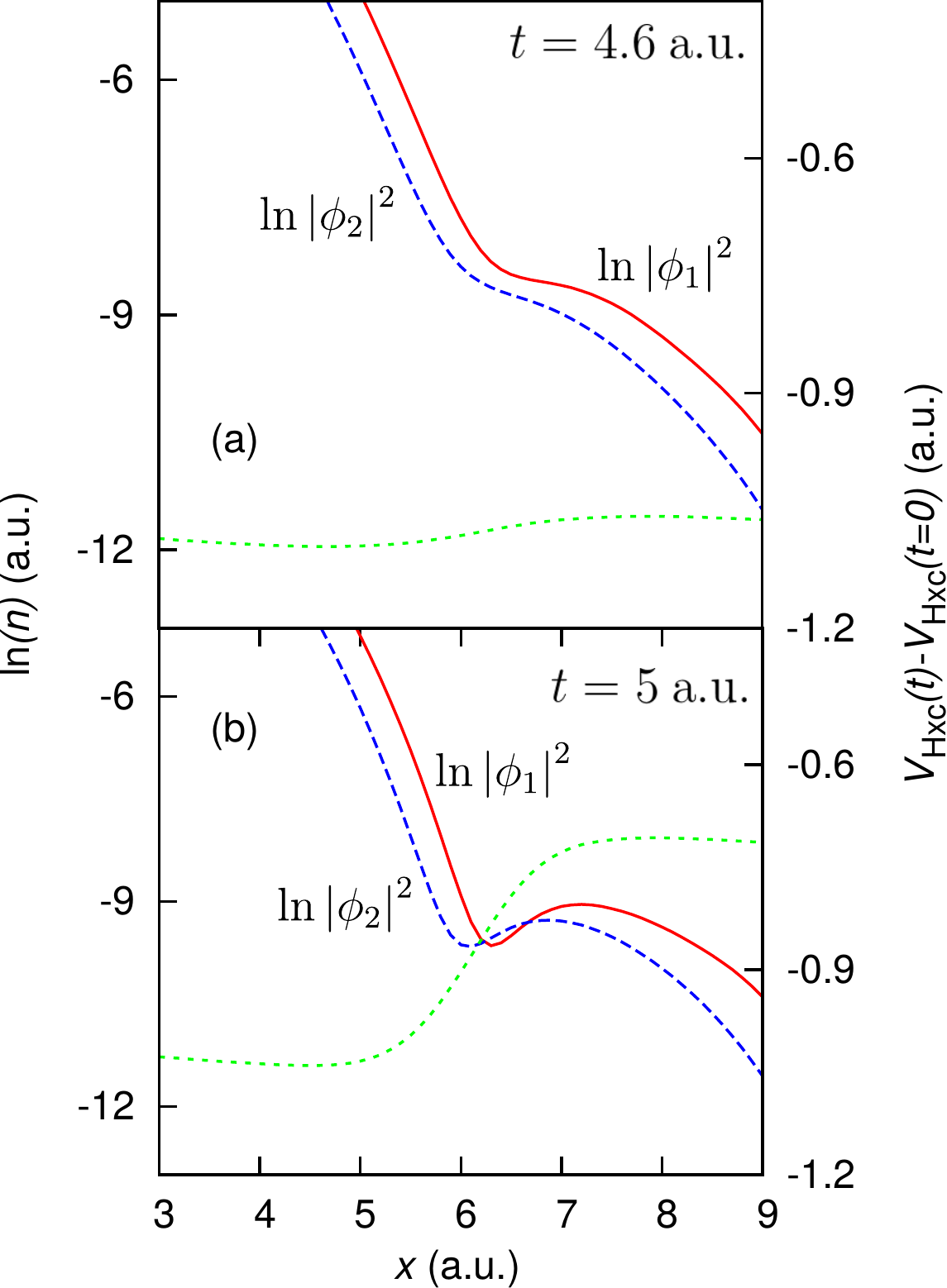}
\caption{(color online) System 3 (dynamic double-well) --- Two electrons in an asymmetric double-well external potential, where a perturbing field ($0.1\left | x \right |$) pushes the electrons together for $t \geq 0$. (a) The natural logarithm of the single-particle KS densities (ground-state -- solid red, first excited-state -- dashed blue), with the time-dependent part of the Hxc potential (short-dashed green) at $t=4.6$ a.u. (b) The same as (a) but at $t=5$ a.u.\ and the single-particle KS densities have crossed, causing a time-dependent step to form in the Hxc potential at the point where the densities cross.}
\label{TD_steps}
\end{figure}

We look at the single-particle time-dependent KS densities for two electrons in an asymmetric double-well external potential\footnote{see Supplemental Material}, where for $t \geq 0$ a perturbing field ($0.1\left | x \right |$) pushes the electrons together (System 3)\repnote, and find that the dynamic steps also occur at the points where the individual KS densities cross, showing that, to some degree, the dynamic steps occur as a result of this phenomenon; Fig.~\ref{TD_steps}. However, this concept is less well defined for the time-dependent case, as the idea of a well defined ionization energy is no longer applicable.

The step here does correspond to a peak in the velocity field (current density divided by electron density), which in turn forms as a result of a minimum in the electron density, as in Ref.~\onlinecite{PhysRevB.88.241102}. We find that in the system studied in Ref.~\onlinecite{PhysRevB.88.241102} there are density minima, and thus peaks in the velocity field, that do \textit{not} correspond to steps in the time-dependent xc potential. We have confirmed that this is because these density minima do not correspond to KS single-particle densities crossing. 

Thus, the question remains; why do the KS single-particle densities seem to always cross at density minima? For dynamic finite systems interference `ripples' in the density are likely to occur\cite{PhysRevB.88.241102}, hence if an orbital density develops an extremum, there is an enhanced likelihood of it crossing an adjacent orbital density. Thus, minima in the dynamic electron density may also serve to indicate where steps will form. However, as the energy levels are not well defined in the dynamic regime the magnitude of the step may vary from that given by Eq.~\ref{step_height}. But, if the system is in the adiabatic limit then our arguments for the ground-state steps would approximately apply for the time-dependent system. 

\subsection{The role of density minima for ground-state systems}
\label{relationship}

A turning point often occurs when the dominant contribution shifts from one electron to another. Thus a density minimum is likely to correspond to a change in the local effective ionization energy and/or a crossover of the single-particle KS densities. Hence, in our calculations we observe that density minima are usually good indicators of where steps will form. Next we will show that steps do \textit{not} form at \textit{all} density minima, as some density minima can not correspond to a change in the local effective ionization energy or this concept is not well defined. However, we demonstrate below how certain density minima, which also represent the interface between localized electrons, are indicators for where in the electron density steps will form for ground-state systems.

Consider a subsystem where the majority of the electron density corresponds to one strongly localized electron. If there is a minimum in the density within the subsystem it cannot correspond to a change in the local effective ionization energy, because there is only one occupied energy state. Thus, there can never be an overall step in $v_\mathrm{xc}$ for a minimum within a subsystem consisting of one electron. This then shows that not all density minima correspond to steps in the xc potential. Yet, the question remains; which density minima will give rise to steps? 

\begin{figure}[htbp]
  \centering
  \includegraphics[width=1.0\linewidth]{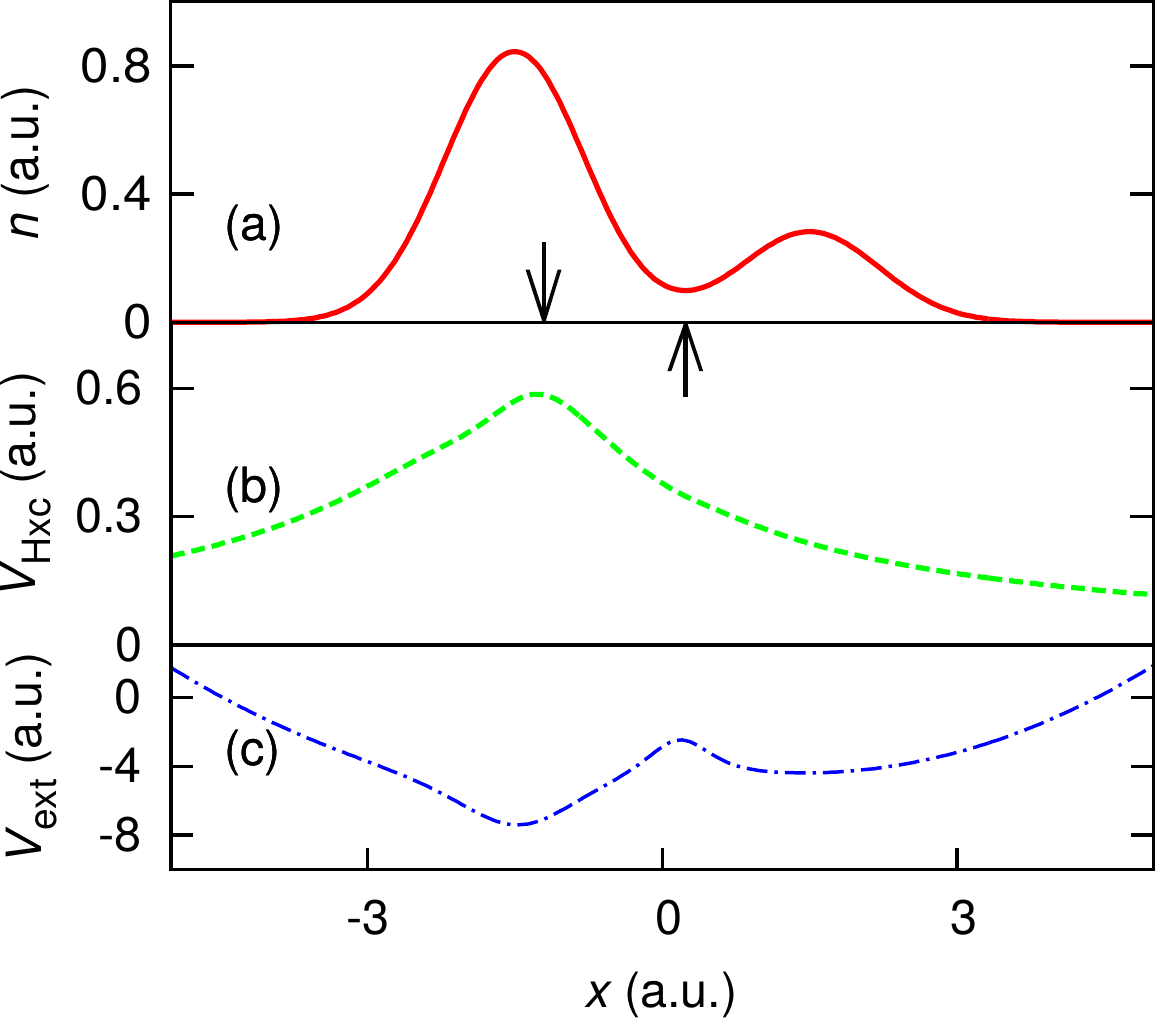}
\caption{(color online) System 5 (crafted external potential) --- (a) The electron density for two interacting electrons (solid red) in a potential crafted such that the IEP (defined by the condition that the electron density to the left of the point integrates to exactly one electron) is distinctly different from the density minimum. The downward facing arrow indicates the IEP at $x \sim -1.2$ a.u., and the upward facing arrow indicates the density minimum at $x \sim 0.24$ a.u. (b) The Hartree exchange-correlation potential (dotted green): the predominant feature of the potential -- not a step -- is at the IEP. (c) The external potential (dashed-dotted blue).}
\label{INT_IEP}
\end{figure}

In systems containing well-separated subsystems, the local effective ionization energy is well defined near the boundary of each subsystem, but can change from one value to another as the boundary is crossed. If the number of electrons in this subsystem integrates to an integer (which is usual for localized systems), we can define the \textit{integer electron point} (IEP) as an indicator of this boundary, and hence of where a step may form. (We note that as a given subsystem may contain several, localized electrons, features in $v_\mathrm{xc}$ within the subsystem may correspond to IEPs due to changes in the local effective ionization energy and/or crossing single-particle KS densities. However the possible delocalization due to the electrons being confined within the subsystem may cause these features to be unrecognisable as steps; see Sections~\ref{effect of delocalization} and \ref{superposition}, and Fig.~\ref{INT_IEP}.) Therefore, if, in 1D, the density minima ($a$ and $b$) satisfy $\int_{a}^{b}n(x)dx=N$, where $N$ is an integer, those density minima are good indicators of where steps (or other features) may form, provided that the IEPs and density minima tend to coincide (which we observe them to). We show below how the Coulomb repulsion and the degree of localization in the system are responsible for density minima and IEPs being at approximately the same point. We note that in the time-dependent regime (as observed above and in Ref.~\onlinecite{PhysRevB.88.241102}), owing to energy levels being less well defined, the IEP is \textit{not} an indicator of a density minimum that may correspond to a step. 

To explore the relationship between density minima and integer electron points (IEPs) in the ground state, we examine how a system may be split into subsystems. With a sufficient degree of localization for all electrons in a system, IEPs indicate the crossover from one electron to the next. In the limit of complete electron localization, the IEPs are definite intersections between the electrons, hence giving a clear boundary between the subsystems. As the electrons delocalize, some of the on-site electron spreads into the neighboring sites. This delocalization, and the effect it has on the shape of steps in the xc potential at the IEPs, is studied below. 

\begin{figure}[htbp]
  \centering
  \includegraphics[width=1.0\linewidth]{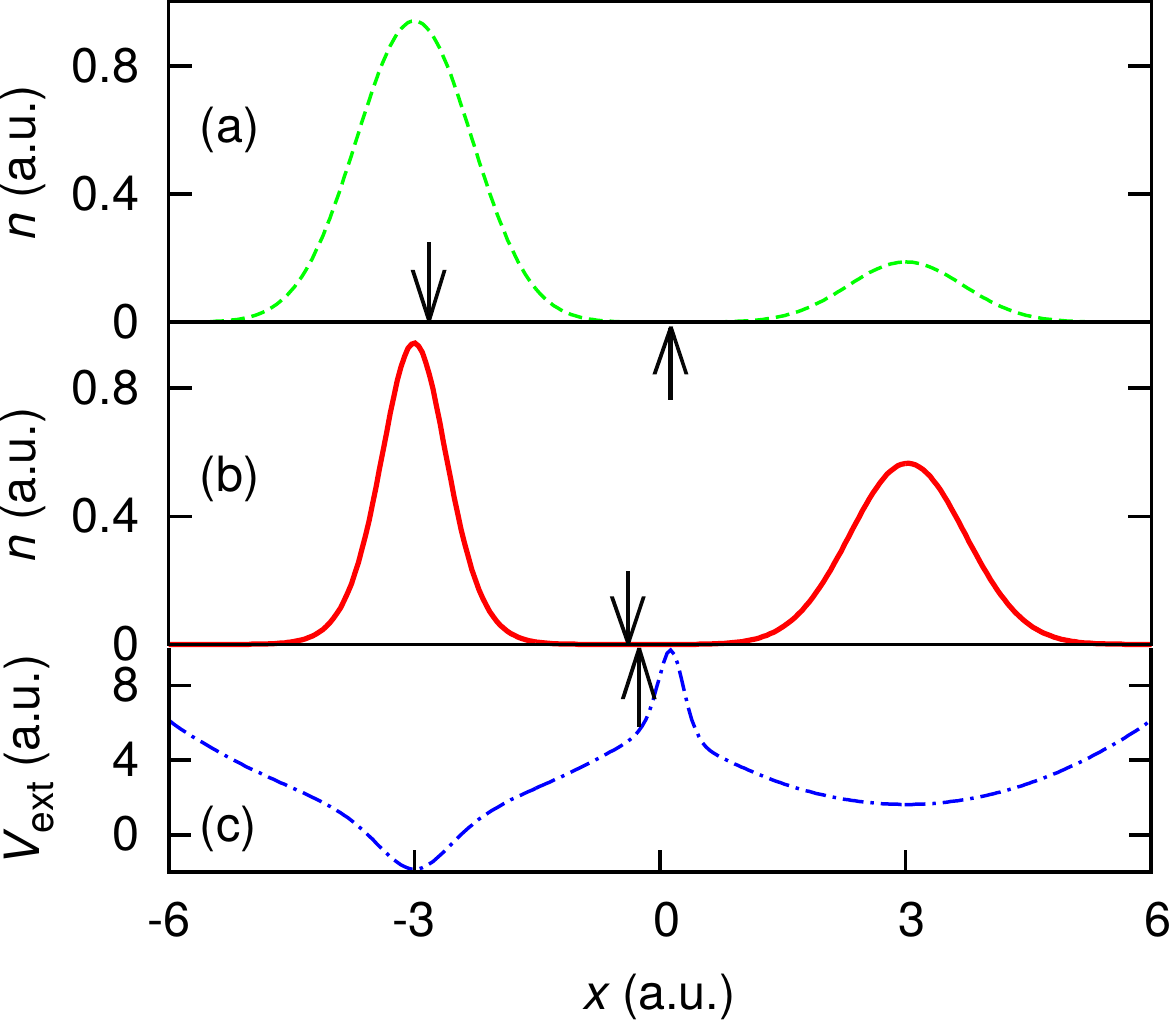}
\caption{(color online) System 4 (crafted external potential) --- (a) The non-interacting electron density (dashed green) for two electrons in the external potential of (c). The IEP (see text) is shown by the downward facing arrow at $x \sim -2.8$ a.u.\ and the density minimum is by the upward facing arrow at $x \sim 0.13$ a.u. (b) The interacting electron density (solid red) for two electrons in the external potential of (c). Again the IEP is shown by the upward facing arrow at $x \sim -0.39$ a.u.\ and the density minimum by the downward facing arrow at $x \sim -0.26$ a.u. The interaction acts to draw the IEP and density minimum together. (c) Shows the external potential for this system. This potential has been crafted so that, for noninteracting electrons, the density minimum and IEP are very different.}
\label{IEP}
\end{figure}

We observe in calculations of electron densities that an IEP typically occurs approximately at a minimum in the electron density. To show that the Coulomb repulsion is largely responsible for this phenomenon, we introduce a 2-electron system (System 4), where the IEP and density minimum are designed to be significantly different for \textit{non-interacting} electrons; see Fig.~\ref{IEP}(a). With two \textit{interacting} electrons in the \textit{same external potential} we observe the IEP and the density minimum tending to the same point; see Fig.~\ref{IEP}(b).

To understand this, we imagine artificially increasing the interaction strength between the electrons: the likelihood of finding the left electron in the right subsystem, i.e.\ to the right of the IEP, and vice versa, reduces owing to the electron repelling the other from its vicinity. Thus, the electrons localize and the density at the IEP tends to zero. For a non-negative quantity, such as the electron density, any zero point must correspond to a minimum.

For the physical interaction strength, it is possible for a system (e.g.\ System 5 below) to have an IEP that does \textit{not} correspond to a minimum in the density of \textit{interacting} electrons. However, achieving this requires a carefully crafted external potential which causes the appropriate degree of delocalization. 

Figure~\ref{INT_IEP} shows that the predominant feature in the Hartree exchange-correlation (Hxc) potential ($v_\mathrm{H}+v_\mathrm{xc}$) forms at the IEP, however there is no step as the local effective ionization energy does not have a well defined value on each side of the feature -- a characteristic of the exact functional shared by the SOA in more delocalized systems such as this one.

To summarize, a change in the local effective ionization energy is required for a step to form -- usually indicated by a density minimum corresponding to an IEP. The IEP and density minimum will be at approximately the same point in the electron density owing to the degree of localization in the system coupled with the Coulomb repulsion. Future improved density functionals may exploit this approximate functional relationship to include features of the exact KS potential examined above.

\section{Sharpness of steps: effect of delocalization}
\label{effect of delocalization}

Considering how the step forms, it is apparent that the more abrupt the switch between dominant KS orbitals (correlated with localization), and between local effective ionization energies, the sharper the step will be. Therefore, next we test what happens to the shape of a step as the region of delocalization increases. (We note that the step forms in the region of highest delocalization, which corresponds to the interface between the electrons\footnote{We have looked at the electron localization function (ELF)\cite{dobson1991interpretation} for this system, and found that the electrons are extremely localized towards the edges. As one approaches the `interface' between the two electrons strong delocalization occurs.}.) Tempel \textit{at al.} considered a singlet case where two potentials were seperated, and the effect on the step was observed\cite{tempel2009revisiting}. Their findings are in agreement with our concept of the local effective ionization energy. They find that as the molecule dissociates the step becomes clear as the separation increases, i.e.\ as the local effective ionization energy becomes well defined (i.e.\ $\tilde{I}(x)\rightarrow I$). 

\begin{figure}[htbp]
  \centering
  \includegraphics[width=1.0\linewidth]{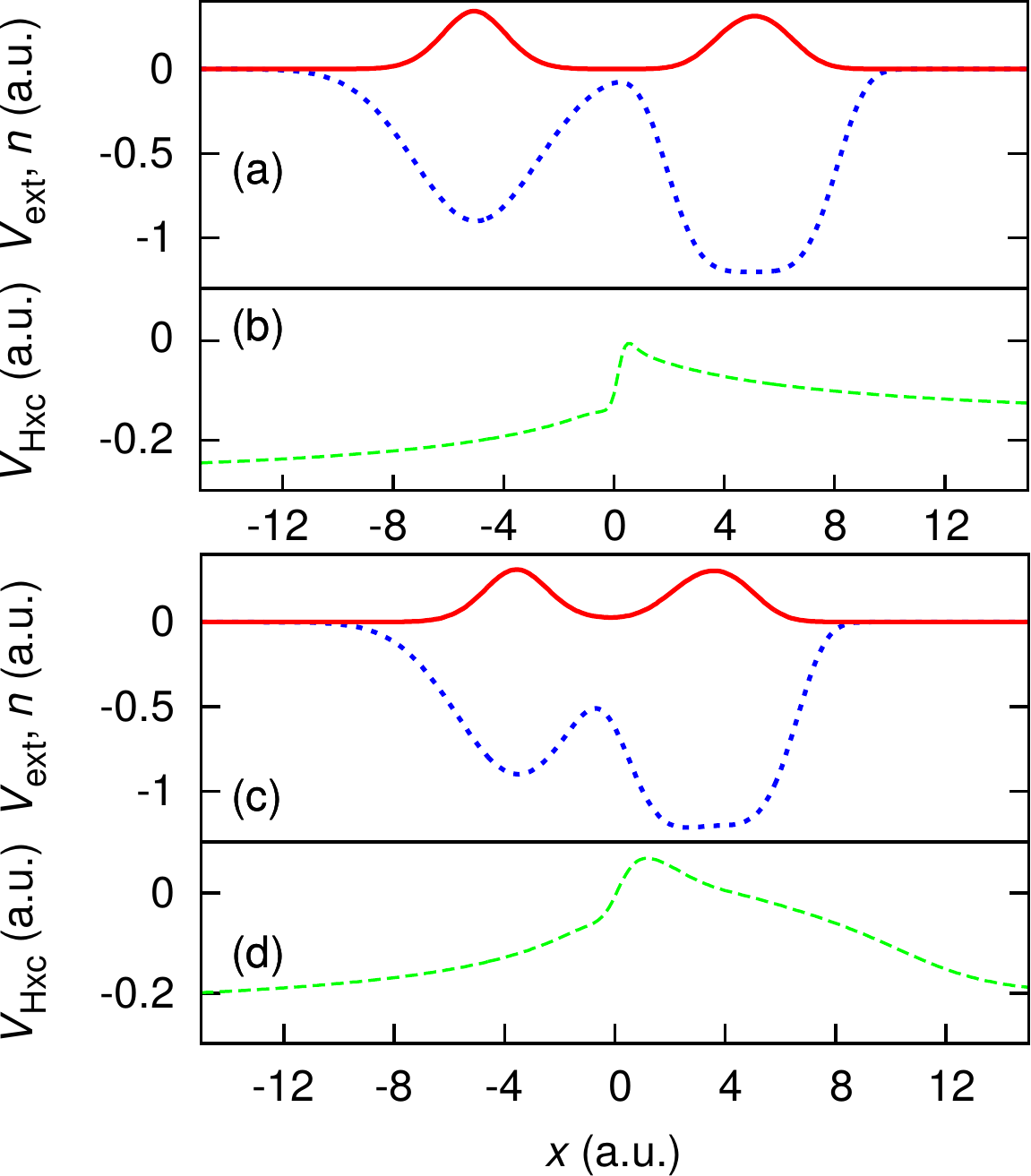}
\caption{(color online) System 6 (increase separation of wells) --- (a) The external potential (dotted blue) and electron density (solid red). (b) The Hartree exchange-correlation (Hxc) potential has a step; this ensures that both KS electron occupy just one well each. (c) The external potential and electron density for System 6$'$. The system is that of System 6, except that the wells have been brought closer together. (d) The Hxc potential for System 6$'$ shows a step, like that of System 6, but because the delocalization is stronger the step is less sharp.}
\label{system1_loc}
\end{figure}

We introduce another system (System 6), which has the usual form: two spinless electrons in an asymmetric double well\repnote. Figure~\ref{system1_loc}(b) shows the Hxc potential for System 6 -- note the sharp step.

Figure~\ref{system1_loc} shows that as the localization decreases the `sharpness' of the step decreases also. This observation is in agreement with our above analysis -- sharp steps cannot form in regions where there is not a well-defined difference in ionization energy. Note the second, very diffuse step about $x\sim11$ a.u.\ where the KS single-particle densities cross once more; see Fig.~\ref{system1_loc}(d). Here the effect is to counteract the step between the electrons so that there is no net step. (In Fig.~\ref{system1_loc}(b) the system is not large enough for the KS densities to cross twice, hence there is only one step.)

We apply the Hartree-Fock (HF) approximation to System 6 as a means of determining the role that exchange plays in these systems. We reverse-engineer the HF electron density using iDEA to find the local potential which describes the density (HF-KS potential). In this way we can compare the steps of the HF-KS potential to those of the exact KS potential. For some systems -- where the KS HOMO and LUMO are distinctly different -- we observe the HF-KS potential to have a step which is almost perfect, as for System 6. Whereas for systems where the KS HOMO and LUMO energies are close, correlation is stronger, and the HF-KS potential's step (and other features) are less accurate. Thus both exchange and correlation may be important in determining the properties of the steps.

\section{Bumps and other superpositions of steps}
\label{superposition}

In the following model systems we demonstrate that the steps in $v_\mathrm{xc}$ in symmetric systems in effect coalesce to form `bumps' in the potential, for systems with some degree of delocalization. 

We demonstrate this by studying two examples -- one time-dependent and one ground-state -- each comprised of three systems (A, B and C). The external potential for the third System (C), in each case, is given by $v_\mathrm{ext}^C=\tfrac{1}{2}\left(v_\mathrm{ext}^A+v_\mathrm{ext}^B\right)$. From this we can find the relationship between the KS potentials for the three systems. We write (to first order)\footnote{To condense the notation, we use implicit matrix-style spatial arguments and integration as is standard in linear response theory.}
\begin{equation}
v_\mathrm{KS}^C = v_\mathrm{KS}^A + \frac{\delta v_\mathrm{KS}}{\delta v_\mathrm{ext}} \left ( v_\mathrm{ext}^C-v_\mathrm{ext}^A  \right ) \nonumber
\end{equation}
and likewise with $A$ replaced by $B$. If we add the two together and divide by two we get 
\begin{equation}
v_\mathrm{KS}^C =\frac{1}{2} \left [v_\mathrm{KS}^A + v_\mathrm{KS}^B + \frac{\delta v_\mathrm{KS}}{\delta v_\mathrm{ext}} \left (2v_\mathrm{ext}^C-v_\mathrm{ext}^A -v_\mathrm{ext}^B  \right )  \right] \nonumber
\end{equation}
thus, provided the systems are sufficiently similar in character to have similar response functions 
\begin{equation}
\left(\frac{\delta v_\mathrm{KS}}{\delta v_{ext}}\right)_C \approx \left( \frac{\delta v_\mathrm{KS}}{\delta v_{ext}} \right)_B \approx \left( \frac{\delta v_\mathrm{KS}}{\delta v_{ext}} \right)_A,
\label{LR}
\end{equation}
we can infer that
\begin{equation}
v_\mathrm{KS}^C \approx  \tfrac{1}{2}\left(v_\mathrm{KS}^A+v_\mathrm{KS}^B\right).
\label{KSsum}
\end{equation}
In the present context the bump potential of system C is the sum of two oppositely-stepped potentials A and B.

\subsection{Ground-state example}

We study three systems to demonstrate, using the above linearity, how positive and negative steps may manifest in a symmetric system as a bump. The bump we observe is very similar in character to that of Ref.~\onlinecite{PhysRevA.49.2421}, where a `peak' in the xc potential arises between atomic shells. Reference~\onlinecite{PhysRevA.54.1957} describes peaks/bumps forming with steps for molecular systems like our own. We show below how steps and bumps both manifest through the superposition of steps in the xc potential. Reference~\onlinecite{PhysRevA.83.062512} also observed a peak in the exact exchange potential at `the crossover point of orbital domination.'

\begin{figure}[htbp]
  \centering
  \includegraphics[width=0.9\linewidth]{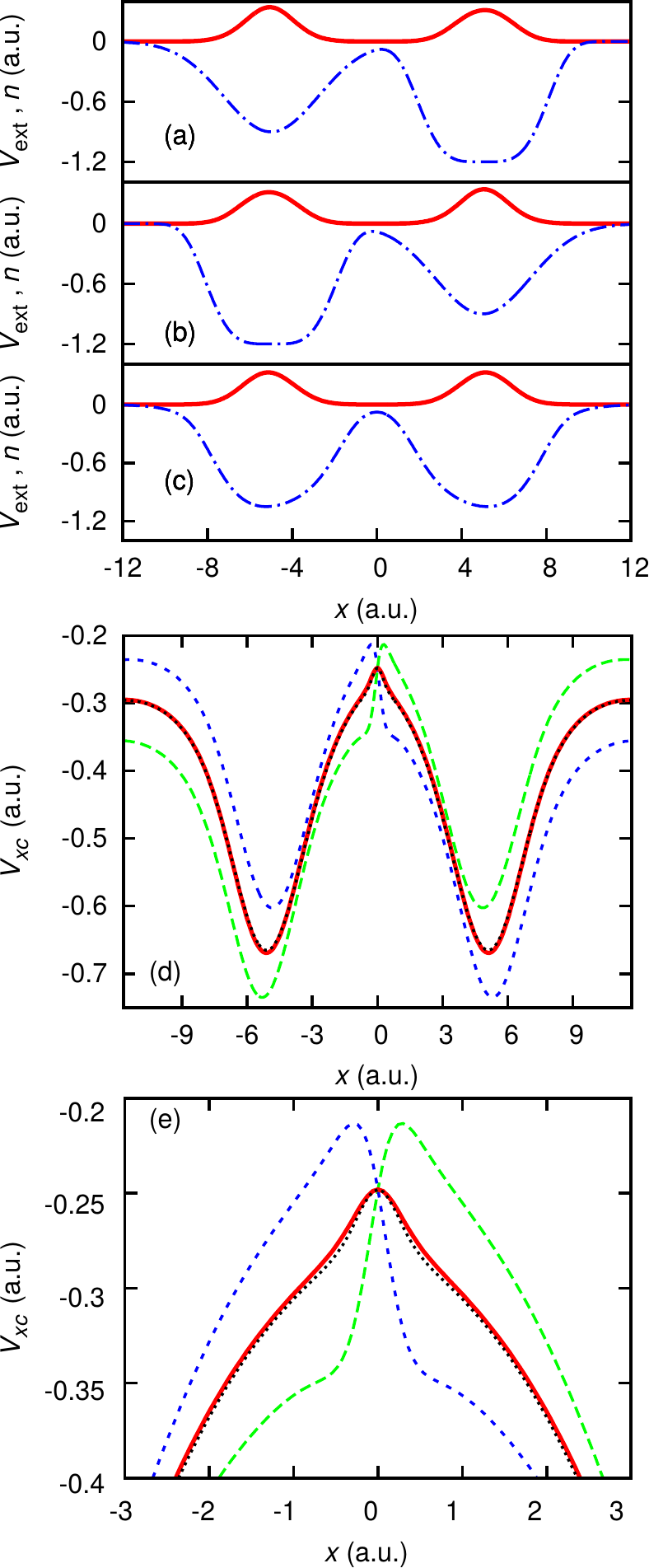}
\caption{(color online) System 7A, 7B and 7C --- (a) The external potential (dotted-dashed blue) and the electron density (solid red) for System 7A. (b) The same for System 7B. (c) The external potential (dotted-dashed blue), defined by averaging the external potentials of 7A and 7B, with the electron density (solid red). (d) The xc potential ($v_\mathrm{xc}^A$) for System 7A (dotted green), with the xc potential ($v_\mathrm{xc}^B$) for System 7B (dashed blue). The xc potential for System 7C (solid red) is compared against $\tfrac{1}{2}(v_\mathrm{xc}^A+v_\mathrm{xc}^B)$ (short-dashed black). We note the good agreement between the two, and how well the bump in the potential is reproduced by the superposition of steps. (e) is a close up of the bump and steps in (d).}
\label{system1}
\end{figure}

System 7A (Fig.~\ref{system1}(a)) is the usual two spinless electrons in an asymmetric external potential\repnote designed to give a step, System 7B (Fig.~\ref{system1}(b)) is the same as System 7A but reflected about $x=0$ (explained below), and the symmetric System 7C (Fig.~\ref{system1}(c)) is the superposition of 7A and 7B (as described above). 

We choose our second system (System 7B) to be the mirror image of System 7A, so that 7C is symmetric. Finally, we construct System 7C from System 7A and 7B (as stated above). The density minimum is aligned at $x=0$ in all three systems. As System 7C is symmetric, no overall steps can form in the exact xc potential of C; instead a bump forms at the density minimum; see Fig.~\ref{system1}(d) and (e). This bump acts to `push' the electrons apart, recreating the effect of the Coulomb repulsion. Figure~\ref{system1}(d) shows the xc potential given by Eq.~\ref{KSsum} as well as the exact xc potential for Systems 7C, 7B and 7A. We observe the precision with which the xc potential of System 7C is replicated by the superposition of steps, as well as the self-interaction correction either side of this central feature. This accuracy is due to Eq.~\ref{KSsum} holding well (in itself a striking result). We have also shown that the symmetric bump feature in the exact xc potential of System 7C can be thought of as the sum of positive and negative steps; see Fig.~\ref{system1}(e) for close up.

We stress that Systems 7A and 7B satisfy the requirement that their differences from System 7C may be described within a linear-response framework (Eq.~\ref{LR}). In this sense, there are several sets of systems which would demonstrate the above superposition of steps to form a bump. We also point out that had System 7A not corresponded to the reflection of 7B about $x=0$, then 7C could be asymmetric and hence may have an overall step. We have simulated this scenario and found that two differently sized steps superimpose to give a step-and-peak combination for System C, reminiscent of the step and peak of Fig.~\ref{Almbladh-von_Barth}(c). 

\subsection{Time-dependent example}

\begin{figure}[htbp]
  \centering
  \includegraphics[width=0.9\linewidth]{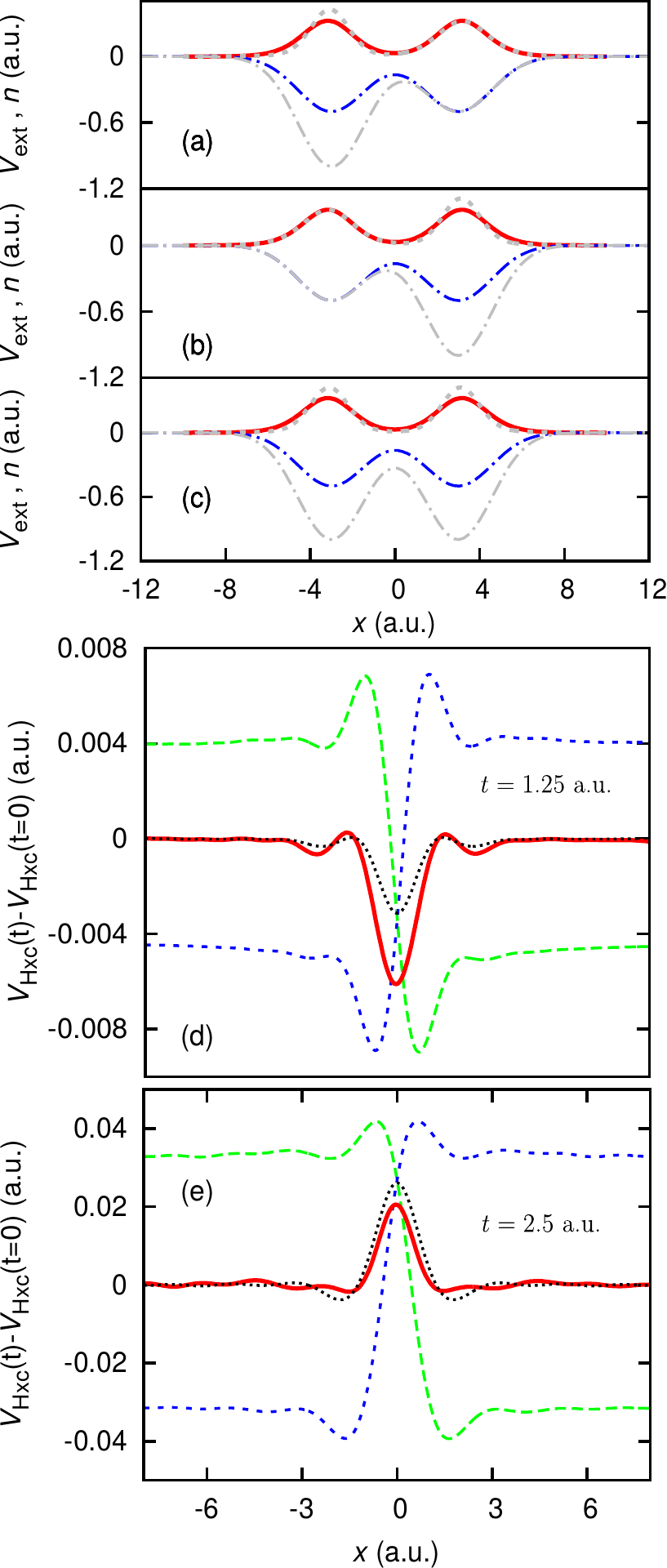}
\caption{(color online) System 8A, 8B and 8C --- (a) The external potential (dotted-dashed blue) and the electron density (solid red) for System 8A at $t=0$. The gray lines indicate the perturbed potential and the electron density at $t=2.5$ a.u. (b) The same for System 8B. (c) The external potential (dotted-dashed blue), defined by averaging the external potentials of 8A and 8B (as for the perturbed potential shown in gray), with the electron density (solid red) and at $t=2.5$ a.u.\ (gray). (d) The dynamic part of the Hxc potential ($v_\mathrm{Hxc}^A(t)-v_\mathrm{Hxc}^A(t=0)$) for System 8A (dotted green), with the same potential for System 8B (dashed blue). The same potential for System 8C (solid red) is compared against the averaged potential (short-dashed black) at $t=1.25$ a.u. We note the good agreement between the two, and how well the `dip' in the potential is reproduced by the superposition of steps. (e) The same graph for $t=2.5$ a.u., the dip has now become a bump.}
\label{TD_superposition}
\end{figure}

We extend this concept of superimposing steps to dynamic systems. We once again consider three systems: the first (System 8A), a symmetric double well\repnote in its ground state, designed such that, for $t \geq 0$, a dynamic steps grows (Fig.~\ref{TD_superposition}(a)); the second (System 8B) the mirror image of the first (Fig.~\ref{TD_superposition}(b)), and the third (System 8C) is symmetric (Fig.~\ref{TD_superposition}(c)). Once again we align the origins of the three systems at the density minima. 

System 8A, in the ground-state, is comprised of two electrons in a double well. At $t=0$ we apply a perturbing field which excites the left electron by increasing the depth of the left well allowing the left electron to explore excited states -- a dynamic step grows at the density minimum as a result. System 8B is the same, but reflected about $x=0$. And System 8C (defined in the same way as the ground-state example) is symmetric, so both electrons explore excited states. As two dynamic steps form, they correctly superimpose at all times to create a feature which oscillates between a bump and a dip; see Fig.~\ref{TD_superposition}(d) and (e). 

\section{Conclusions}

Knowledge of how the positions, magnitudes and shape of xc potential steps depend on features of the density, such as the locations of minima and the local ionization energies, provides an important basis for the construction of improved density functionals.

We have introduced the concept of the `local effective ionization energy' which applies in regions far from an atom, where the ionization energy associated with the a single electron is well defined. At an interface between localized electrons the local effective ionization energy can change; when this happens over a short range it gives rise to a step in the exchange-correlation (xc) potential with a magnitude equal to the difference in ionization energies. For systems with more than one occupied Kohn-Sham (KS) orbital, the analog of this effect, in the KS picture, arises from the crossover from one localized single-particle KS density to another. At this point, the dominant contributing KS orbital to the overall electron density changes. This effect also gives rise to a step, even in the time-dependent regime.

We build on the thought experiment of Almbladh and von Barth\cite{almbladh1985density} by considering the above effects far from a pair of separated atoms. One step forms between the atoms and corrects the number of electrons on each atom. The second step, far from the molecule, also corresponds to a change in the local effective ionization energy and a crossover of the single-particle KS densities.

We also derive from this fundamental understanding of steps a practical means of approximating where steps in the electron density will form, and provide a range for the step height to ensure accurate electron densities. We find that steps usually require a minimum in the electron density in order to form; however not all density minima yield steps, as the minimum must correspond to the interface of at least two localized electrons. By integrating over a region of electrons which is localized relative to other electrons in the system, we can define an integer electron point, i.e.\ the point in the density where this crossover occurs thus a step can form. Localization indicates which systems will require steps, while the degree of localization affects the shape of the steps. 
Further, linear-response theory shows how various other features in the KS potential can be interpreted as the superposition of steps, even in time-dependent systems. 

\section{Acknowledgements}

We acknowledge assistance from Richard Lynn with Hartree-Fock calculations, from Thomas Durrant with characterization of electron localization, and funding from EPSRC.

\newpage
\bibliography{Bibtex}

\end{document}


\title{Origin of static and dynamic steps in exact Kohn-Sham potentials (Supplemental Material)}
\date{\today}
\author{M.\ J.\ P.\ Hodgson}
\author{J.\ D.\ Ramsden}
\author{R.\ W.\ Godby}
\affiliation{Department of Physics, University of York and European Theoretical Spectroscopy Facility, Heslington, York YO10 5DD, United Kingdom} 

\maketitle

\section{The Almbladh-von Barth thought experiment} 
The external potential is (we use Hartree atomic units)
\begin{equation}
v_{\mathrm{ext}} = - \tfrac{6}{5}e^{-\frac{1}{125}(x-a)^4} - \tfrac{9}{10} e^{-\frac{1}{10}(x+a)^2},
\label{eq1}
\end{equation}
where $a=10$ a.u. For this system converged results are obtained with $\delta x = 0.1$ a.u.

\section{System 1 (two spinless electrons in an asymmetric double well)} 
The external potential is the same as the Almbladh-von Barth thought experiment, with a grid spacing $\delta x = 0.1$ a.u. 

\section{System 2 (molecule)}
The external potential is of the same form as Eq.~\ref{eq1} with $a=4$ a.u. The wells are brought together so that the second crossover of the single-particle KS densities occurs while the density is within machine precision. The grid spacing $\delta x = \tfrac{4}{30}$ a.u. 

\section{System 3 (dynamic double-well)}
The ground-state external potential is 
\begin{equation}
v_{\mathrm{ext}} = - \tfrac{13}{20}e^{-\frac{1}{2}(x-3)^2} - \tfrac{1}{2} e^{-\frac{1}{10}(x+3)^2}.
\label{eq2}
\end{equation}
For $t \geq 0$ a perturbing field ($0.1\left | x \right |$) pushes the electrons together. The grid spacings are $\delta x = 0.1$ a.u.\ and $\delta t = 1\times 10^{-4}$ a.u.

\section{System 4}
The external potential was optimized in order to give a non-interacting density where the integer electron point and density minimum are distinctly different, and hence has no analytical form. 

\section{System 5}
The external potential was optimized in order to give an interacting density where the integer electron point and density minimum are distinctly different, and hence has no analytical form. 

\section{System 6}
The first system (System 6) has an external potential of the form of Eq.~\ref{eq1} with $a=5$ a.u. The second system (System 6$'$) has an external potential of the form of Eq.~\ref{eq1} with $a=3.5$ a.u. The wells have been brought closer together so that the electrons are more delocalized in the region of the step.

\section{Systems 7A, 7B and 7C (ground-state)}
The external potential of 7A is of the same form as Eq.~\ref{eq1}, with $a=5$ a.u. The external potential of 7B is the mirror of 7A, i.e.\ $a \rightarrow -a$. And the third system's (System 7C) external potential is given by $v_\mathrm{ext}^C=\tfrac{1}{2}\left(v_\mathrm{ext}^A+v_\mathrm{ext}^B\right)$. For these three systems converged results were found with $\delta x = 0.05$ a.u.

\section{Systems 8A, 8B and 8C (time-dependent)}
The ground-state of System 8A ($t < 0$) has an external potential of the form
\begin{equation}
v_{\mathrm{ext}} = - \tfrac{1}{2}e^{-\frac{1}{5}(x+3)^2} - \tfrac{1}{2} e^{-\frac{1}{5}(x-3)^2}.
\label{eq3}
\end{equation}
For $t \geq 0$ the external potential is 
\begin{equation}
v_{\mathrm{ext}}(t \geq 0) = - e^{-\frac{1}{5}(x+3)^2} - \tfrac{1}{2} e^{-\frac{1}{5}(x-3)^2}.
\label{eq3}
\end{equation}
Hence the left well is reduced in depth. This causes the left electron to explore excited energy states.

System 8B is the mirror of System 8A, and again System 8C's external potential is given by $v_\mathrm{ext}^C(t)=\tfrac{1}{2}\left(v_\mathrm{ext}^A(t)+v_\mathrm{ext}^B(t)\right)$. For these three systems converged results were found with $\delta x = 0.05$ a.u.\ and $\delta t = 5\times 10^{-5}$ a.u.